\RequirePackage{fix-cm}
\documentclass[twocolumn,epjc3]{svjour3}
\smartqed  % flush right qed marks, e.g. at end of proof
\RequirePackage{graphicx}
\journalname{Eur. Phys. J. C}
\usepackage{latexsym}
\usepackage{graphics}
\usepackage{bm}
\usepackage{dcolumn}
\usepackage[usenames,dvipsnames]{color}
\usepackage{eufrak}
\usepackage{lscape}
\usepackage{slashed}

\begin{document}
\title{Study of $\Lambda_b\rightarrow~ \Lambda l^+l^-$ and $\Lambda_b\rightarrow p l \bar{\nu}$ decays in the Bethe-Salpeter equation approach}


\author{Y. Liu\thanksref{e1,addr1}
        \and
        L.-L. Liu\thanksref{e2,addr1}
        \and
        X.-H. Guo\thanksref{e3,addr1} %etc.
}

%\thankstext[$\star$]{t1}{Thanks to the title}
\thankstext{e1}{e-mail: yingliubnu@gmail.com}
\thankstext{e2}{e-mail: liu06\_04@mail.bnu.edu.cn}
\thankstext{e3}{e-mail: Corresponding author. xhguo@bnu.edu.cn}

\institute{College of Nuclear Science and Technology,
Beijing Normal University, Beijing
100875, People's Republic of China\label{addr1}
%          \and
%          Second Address, Street, City, Country\label{addr1}
%          \and
%          \emph{Present Address:} Street, City, Country\label{addr1}
}

\date{Received: date / Accepted: date}

\maketitle

\begin{abstract}
In our previous work, based on the $SU(6)$ spin-flavor wave function, we regard $\Lambda$ and $p$ as composed of different quark-diquark configurations and established the Bethe-Salpeter (BS) equations of configurations for quark and scalar diquark. In our present work, we apply this model to calculate the form factors of the semileptonic transitions $\Lambda_b\rightarrow\Lambda l^+l^-$ $(l=\mu,e,\tau)$ and $\Lambda_b\rightarrow p l\bar{\nu}$ within the Standard Model (SM). The decay $\Lambda_b\rightarrow\Lambda \mu^+\mu^-$ is especially interesting since it has been measured in CDF and LHCb Collaborations and this rare decay is very sensitive to new physics effects. The decay $\Lambda_b\rightarrow p l\bar{\nu}$ is a promising mode for the measurement of the Cabibbo-Kobayashi-Maskawa matrix element $|V_{ub}|$ at the Large Hadron Collider. In our calculations, depending on the ranges of the parameters in the model including the diquark mass and the interaction strength between the quark and the diquark in the kernel of the BS equation, we find that the branching ratio of $\Lambda_b\rightarrow\Lambda\mu^+\mu^-$ in our model is consistent with the experimental data and the current experimental results from LHCb agree with the differential branching ratio of $\Lambda_b\rightarrow\Lambda\mu^+\mu^-$ from our calculation except at the lager momentum transfer region. This indicates that there is still room for possible new physics effects. We also give comparisions of the total branching ratios of $\Lambda_b\rightarrow\Lambda l^+l^-$ and $\Lambda_b\rightarrow p l\bar{\nu}$ with those given by other phenomenological methods.
\end{abstract}

\section*{I. Introduction}
In recent years, a lot of experimental progresses have been made in spectroscopy and decays of heavy baryons containing a heavy bottom or charm quark \cite{C1,C2,C3,C4,C5}. Processes associated with the flavor-changing neutral current $b\rightarrow s$ transition have regained much attention since the CLEO measurement of the radiative $b\rightarrow s\gamma$ decay \cite{bs}. Although the experimental measurement of mesonic $b\rightarrow s$ transitions appeared about twenty years ago \cite{meson}, the first observation of the baryonic decay $\Lambda_b\rightarrow\Lambda \mu^+ \mu^-$ was reported in 2011 by the CDF Collaboration \cite{CDF}. A first measurement of the differential and total branching fractions for this rare decay by LHCb was reported in 2013 \cite{LHCb}. The decay $\Lambda_b\rightarrow \Lambda l^+ l^-$$(l=\mu,e,\tau)$ proceeds through electroweak loop diagrams in the Standard Model (SM). Since non-Standard Model particles such as supersymmetric particles \cite{SP} and light dark matter particals \cite{dm} may also participate in these loops, measurement of this decay can be used to search for new physics \cite{NEW}. Furthermore, this channel can be used as a tool in the exact determination of the Cabibbo-Kobayashi-Maskawa (CKM) matrix elements, $V_{tb}$ and $V_{ts}$, and in the study of CP and T violations. On the other hand, the study of the exclusive decay $\Lambda_b\rightarrow p l\bar{\nu}$ is a promising mode for the measurement of the poorly known magnitude of the CKM matrix element $|V_{ub}|$ at the Large Hadron Collider. So far, all the measurements of $|V_{ub}|$ have been from $B$ meson decays and were performed at B factories \cite{Bfac}.

Theoretically, there are some works devoted to the analysis of $\Lambda_b\rightarrow \Lambda l^+l^-$ decays in the SM and in various scenarios of physics beyond the SM \cite{SM1,SM2,SM3,SM4,SM5,SM6,SM7}. In order to use the decay to search for new physics, one should determine the hadronic matrix element $\Lambda_b\rightarrow \Lambda$, which is expressed in twelve form factors. In the heavy quark limit, with the application of the heavy quark effective theory (HEQT) for the $b$ quark, $\Lambda_b\rightarrow\Lambda (p)$ transition can be described by two independent form factors \cite{2ff}. The decrease in the number of form factors greatly simplifies calculations. However, these two form factors contain all soft QCD effects which are difficult to calculate from the first principles. Therefore, one needs to resort to some phenomenological models. The $\Lambda_b\rightarrow \Lambda$ form factors were calculated in quark models \cite{bm,bm1,rm1,rm2,rm3,rm4} and the perturbative QCD approach \cite{pq}. Moreover, the simple pole model was adopted to compute the form factors by Mannel and Rochsiegel \cite{pm}. The authors of Refs. [15-16,29-37] employed the widely applied approach of QCD sum rules to calculate these two form factors. The information on $\Lambda_c\rightarrow\Lambda$ form factors is available from experimental measurement of the semileptonic $\Lambda_c\rightarrow \Lambda e^+\nu_e$ decay \cite{ec,ec2}, and this information was used to constrain the $\Lambda_b\rightarrow\Lambda$ form factors \cite{pm,bm}. Recently, the form factors of the $\Lambda_b \rightarrow\Lambda$ transition were determinated in the first lattice QCD simulation \cite{lattice}. In our previous work, in the ``quark-diquark" model in which a baryon is regarded as a bound state of a quark and a diquark, we established the Bethe-Salpeter (BS) equations for the quark and scalar diquark configurations of $\Lambda$ and $p$ based on $SU(6)$ spin-flavor wave functions \cite{Lb,Lb2}. Then we solved them in the covariant instantaneous approximation with the kernel containing both the scalar confinement and one-gluon-exchange terms \cite{Lb,Lb2,K1,K3,K4,K5}. In the present work, we will apply the BS wave functions of $\Lambda$ and $p$ and those of $\Lambda_b$ which were obtained previously to calculate branching ratios for $\Lambda_b\rightarrow\Lambda l^+l^-$ and $\Lambda_b\rightarrow p l\bar{\nu}$ and compare our results with experiment, lattice data, and results from other phenomenological methods.

The layout of the paper is as follows. in Section II, we will review the basic formalism for the BS equations for $\Lambda$ and $p$. With the aid of HQET, the form factors which are involved in $\Lambda_b\rightarrow\Lambda l^+l^-$ and $\Lambda_b\rightarrow p l\bar{\nu}$ decays will be calculated using the BS wave functions of $\Lambda$, $p$ and those of $\Lambda_b$ which were obtained previously. In Section III we will show calculations and the numerical results of the branching ratios of $\Lambda_b\rightarrow\Lambda l^+l^-$. In Section IV, we will give the branching ratios of $\Lambda_b\rightarrow pl \bar{\nu}$, and will compare with the results from QCD light-cone sum rule. Section V will be deserved for a summary and some discussions.

\section*{II.  Form Factors for $\Lambda_b\rightarrow\Lambda l^+l^-$ and $\Lambda_b\rightarrow p l\bar{\nu}$ }
$\Lambda(p)$ contains three light quarks $u$, $d$, and $s$ $(u)$, in which all the three light quarks play important roles in the dynamics inside the baryon. In general, the parity of a baryon at the ground state is positive. Since the parity of the quark in the baryon is supposed to be positive, the parity of the diquark involved in the ground state baryon should also be positive. Due to Pauli principle, two quarks with the same flavor constitutes an axial-vector diquark and two quarks with different flavors can constitute either a scalar diquark or an axial-vector diquark. Based on the $SU(6)$ wave function of the proton, the proton state can be expanded in the terms of quark-diquark configurations as follows \cite{state}
\begin{eqnarray}
  p^{\uparrow}&=&\frac{1}{3\sqrt{2}}[3u^{\uparrow}(ud)_{0,0} + u^{\uparrow}(ud)_{1,0} - \nonumber\\
  & & \sqrt{2}u^{\downarrow}(ud)_{1,1} - \sqrt{2}d^{\uparrow}(uu)_{1,0}  + 2d^{\downarrow }(uu)_{1,1}],
\end{eqnarray}
\begin{eqnarray}
   p^{\downarrow}&=&\frac{1}{3\sqrt{2}}[3u^{\downarrow}(ud)_{0,0} - u^{\uparrow}(ud)_{1,-1} + \nonumber\\ & &\sqrt{2}u^{\uparrow}(ud)_{1,-1} + \sqrt{2}d^{\downarrow}(uu)_{1,0}  - 2d^{\uparrow }(uu)_{1,-1}].
\end{eqnarray}
In the same way, we can obtain the following forms for $\Lambda$ \cite{state}:
\begin{eqnarray}
  \Lambda^{\uparrow}&=&\frac{1}{2\sqrt{3}}[2s^{\uparrow}(ud)_{0,0} + \sqrt{2}d^{\downarrow}(us)_{1,1}-d^{\uparrow}(us)_{1,0} + \nonumber\\
   & & d^{\uparrow}(us)_{0,0} - \sqrt{2}u^{\downarrow}(ds)_{1,1} + u^{\uparrow}(ds)_{1,0} - \nonumber\\
   & & u^{\uparrow}(ds)_{0,0}]
\end{eqnarray}
\begin{eqnarray}
  \Lambda^{\downarrow}&=&\frac{1}{2\sqrt{3}}[2s^{\downarrow}(ud)_{0,0} - \sqrt{2}d^{\uparrow}(us)_{1,-1} + d^{\downarrow}(us)_{1,0} \nonumber\\
   & &+ d^{\downarrow}(us)_{0,0}+ \sqrt{2}u^{\uparrow}(ds)_{1,-1} -u^{\downarrow}(ds)_{1,0} - \nonumber\\
   & & u^{\downarrow}(ds)_{0,0}].
\end{eqnarray}
\noindent In Eqs. (1-4) the first and the second subscripts correspond to the total spin and the third component of the spin of the diqaurk, respectively. The arrow $\uparrow$ ($\downarrow$) indicates the spin direction of the corresponding baryon is up (down).

$\Lambda_b$ is regarded as a bound state of a $b$ quark and a scalar diquark $(ud)_{0,0}$. In order to calculate the form factors in the transition $\Lambda_b\rightarrow\Lambda$ $(p)$, where the $b$ quark decays into $s$ $(u)$ quark and the scalar diquark behaves as a spectator, one should calculate the BS wave functions of the configurations $s(ud)_{0,0}$ and $u(ud)_{0,0}$ in $\Lambda$ and $p$, respectively.

We define the BS wave function of the system $q(ud)_{0,0}$ $(q=s$ or $u$) as the following:
\begin{eqnarray}
\chi(x_{1},x_{2},P)=\langle 0|T\psi(x_{1})\varphi(x_{2})|P\rangle,
                  % =e^{iPX}\int\frac{d^{4}p}{(2\pi)^{4}}e^{ipx}~\chi_{P}(p),
\end{eqnarray}
\noindent where $\psi(x_{1})$ and $\varphi(x_{2})$ are the field operators of the light quark at position $x_{1}$ and the light scalar diquark at position $x_{2}$, respectively, $P=Mv$ is the momentum of $\Lambda$ or $p$, and $M$ ($v$) is its mass (velocity). Let $m_{q}$ and $m_{D}$ represent the masses of the light quark and the light diquark in the baryon $\Lambda$ or $p$, $\lambda_{1}=\frac{m_{q}}{m_{q}+m_{D}}$, $\lambda_{2}=\frac{m_{D}}{m_{q}+m_{D}}$, and $p$ represent the relative momentum of the two constituents. $X=\lambda_{1}x_{1}+\lambda_{2}x_{2}$ is the coordinate of the center of mass and $x=x_{1}-x_{2}$. The BS wave function in momentum space, $\chi_p(p)$, is related to $\chi(x_{1},x_{2},P)$ through the following equation,

\begin{eqnarray}
\chi(x_{1},x_{2},P)= e^{iPX}\int\frac{d^{4}p}{(2\pi)^{4}}e^{ipx}~\chi_{P}(p).
\end{eqnarray}

The BS equation for the $q(ud)_{0,0}$ system in momentum space can be written as follows:

\begin{eqnarray}
\chi_{P}(p)=S_{F}(p_{1})\int\frac{d^{4}q}{(2\pi)^{4}}K(P,p,q)~\chi_{P}(q)~S_{D}(p_{2}),
\end{eqnarray}
\noindent where $p_{1}=\lambda_{1}P+p$ and $p_{2}=\lambda_{2}P-p$ are the momenta of the light quark $q$ and the light scalar diquark, respectively, $K(P,p,q)$ is the kernel which is defined as the sum of two particle irreducible diagrams, $S_{F}(p_{1})$ and $S_{D}(p_{2})$ are propagators of the light quark with momentum $p_{1}$ and the light diquark with momentum $p_{2}$. Motivated by the potential model, the kernel is given by \cite{Lb2,EK}
\begin{eqnarray}
-iK(P,p,q)=I\otimes IV_{1}(p,q)+{\gamma}_{\mu}\otimes{\Gamma}^{\mu}V_{2}(p,q),
\end{eqnarray}
\noindent where ${\Gamma}^{\mu}=(p_2+q_2)^{\mu}F(Q^{2})$ is the effective vertex of a gluon with two scalar diquarks, $F(Q^{2})$ is introduced to describe the structure of the diquark \cite{Lb2,EK}, $F(Q^{2})=\frac{\alpha_{seff}Q^{2}_{0}}{Q^{2}+Q^{2}_{0}}$, where $Q^{2}_{0}$ is a parameter which freezes $F(Q^{2})$ when $Q^{2}$ is very small. In the high energy region the form factor is proportional to $\frac{1}{Q^{2}}$ which is consistent with perturbative QCD calculations. By analyzing the electromagnetic form factor for the proton, it was found that $Q^{2}_{0}=3.2 $GeV$^{2}$  can lead to consistent results with the experimental data.
 $V_1$ and $V_2$ are the scalar confinement and one-gluon-exchange terms which have the following forms in the covariant instantaneous approximation respectively \cite{Lb,Lb2,K1,K3,K4,K5},
\begin{eqnarray}
\tilde{V}_{1}(p_{t}&&-q_{t})=\frac{8\pi\kappa}{[(p_{t}-q_{t})^{2}+\mu^{2}]^{2}}\nonumber\\
& &-(2\pi)^{3}\delta^{3}(p_{t}-q_{t})\int\frac{d^{3}k}{(2\pi)^{3}}\frac{8\pi\kappa}{(k^{2}+\mu^{2})^{2}},
\end{eqnarray}
\begin{eqnarray}
\tilde{V}_{2}(p_{t}-q_{t})=-\frac{16\pi}{3}\frac{\alpha_{seff}}{(p_{t}-q_{t})^{2}+\mu^{2}},
\end{eqnarray}
\noindent where $p_{t}$ and $q_{t}$ are the transverse projection of the relative momentum along the momentum $P$, which are defined as $p_t^{\mu}=p^{\mu}-v\cdot p v^{\mu}$ and $q_t^{\mu}=q^{\mu}-v\cdot q v^{\mu}$. The second term of $\tilde{V}_1$ is introduced to remove the infrared singularity at the point $p_{t}=q_{t}$, and the small parameter $\mu$ is introduced to avoid the divergence in numerical calculations.
After considering the constraints on $\chi_{P}(p)$ imposed by parity and Lorentz transformations, $\chi_{P}(p)$ can be expressed in terms of two Lorentz-scalar functions, $f_1$ and $f_2$,

\begin{eqnarray}
 \centering
  \chi_{P}(p)=(f_1+\slashed{p_{t}}f_2) u(v,s),
\end{eqnarray}

\noindent where $u(v,s)$ is the Dirac spinor of $\Lambda$ or $p$. Defining $\tilde{f}_{1(2)}(=\int\frac{dp_{l}}{2\pi}f_{1(2)})$, we find these two BS scalar wave functions satisfy the coupled integral equations as follows:
\begin{eqnarray}
  \tilde{f}_{1}(p_{t})&=& -\int\frac{d^{3}p_{t}}{(2\pi)^{3}}\frac{1}{4\omega_{q}\omega_{D}(M-\omega_{q}-\omega_{D})}\nonumber \\
  & &\big[(m_{q}+\omega_{q})(\tilde{V}_{1}+2\omega_{D}\tilde{V}_{2}F(Q^{2}))\tilde{f}_{1}(q_{t})\nonumber \\
 & &-(p_{t}\cdot q_{t}+p_{t}^2)\tilde{V}_{2}F(Q^{2})\tilde{f}_{1}(q_{t})\big]\nonumber \\
 & &  -\int\frac{d^{3}p_{t}}{(2\pi)^{3}}\frac{1}{4\omega_{q}\omega_{D}(M-\omega_{q}-\omega_{D})}\nonumber \\
 & &\big[(\tilde{V}_{1}-2\omega_{D}\tilde{V}_{2}F(Q^2))p_{t}\cdot q_{t}\tilde{f}_{2}(q_{t})\nonumber \\
& &-(m_q+\omega_q)(p_{t}\cdot q_{t}+q_{t}^2)\tilde{V}_2F(Q^2)\tilde{f}_2(q_{t})\big],
\end{eqnarray}
\begin{eqnarray}
   \tilde{f}_{2}(p_{t})&=& -\int\frac{d^{3}p_{t}}{(2\pi)^{3}}\frac{1}{4\omega_{q}\omega_{D}(M-\omega_{q}-\omega_{D})}\nonumber \\
  & &\big[(\tilde{V}_{1}+2\omega_{D}\tilde{V}_{2}F(Q^{2}))\tilde{f}_{1}(q_{t})\nonumber \\
& & -(m_q-\omega_q)\frac{(p_{t}\cdot q_{t}+p_{t}^2)}{p_{t}^2}\tilde{V}_{2}F(Q^{2})\tilde{f}_{1}(q_{t})\big]\nonumber \\
& &
-\int\frac{d^{3}p_{t}}{(2\pi)^{3}}\frac{1}{4\omega_{q}\omega_{D}(M-\omega_{q}-\omega_{D})}\nonumber \\
  & &\big[(m_q-\omega_q)(\tilde{V}_{1}-2\omega_{D}\tilde{V}_{2}F(Q^2))\frac{p_{t}\cdot q_{t}}{p_{t}^2}\tilde{f}_{2}(q_{t})\nonumber \\
& & -(p_{t}\cdot q_{t}+q_{t}^2)\tilde{V}_2F(Q^2)\tilde{f}_2(q_{t})\big],
\end{eqnarray}

where $\omega_D=\sqrt{m^2_D+p^2_t}$ and $\omega_q=\sqrt{m^2_q+p^2_t}$. The BS wave functions $\tilde{f}_{1(2)}$ can be solved numerically by discretizing the integration region $(0,\infty)$ into $n$ pieces ($n$ is chosen to be sufficiently large). The normalization condition for the BS wave function is given in the following after imposing the covariant instantaneous approximation on the kernel \cite{K1,K5,N1}:

\begin{eqnarray}
  i\delta^{i_{1}i_{2}}_{j_{1}j_{2}}\int\frac{d^4qd^4p}{(2\pi)^8}\bar{\chi}_P(p,s)&&
  \bigg[\frac{\partial}{\partial P_0}I_{P}(p,q)^{i_1i_2j_2j_1}\bigg]\nonumber\\
  & & =\delta_{ss'} \chi_P(q,s')
\end{eqnarray}

where $i_{1(2)}$ and $j_{1(2)}$ represent the color indices of the light diquark and the light quark, respectively, $s^{(')}$ is the spin index for the light baryon, $$\delta^{i_1,i_2}_{j_1,j_2}=\delta^{i_1}_{j_1}\delta^{i_2}_{j_2}-\delta^{i_1}_{j_2}\delta^{i_2}_{j_1},$$  $I_{P}^{i_{1}i_{2}j_{2}j_{1}}(p,q)$ is the inverse of the four point propagator defined as follows:

\begin{eqnarray}
  I_P^{i_1i_2j_2j_1}(p,q)&&=\delta^{i_1j_1}\delta^{i_2j_2}(2\pi)^4\delta^4(p-q)S^{(-1)}_{q}(p_1)S^{-1}_D(p_2).\nonumber\\
  &&
\end{eqnarray}

In our calculations, we choose the diquark mass $m_D$ to range from 700 MeV to 800 MeV \cite{rm4}. With this choice for $m_D$, the binding energy $E$ is negative and varies  from -90 MeV to -190MeV. $\kappa$ is chosen to range from 0.02 GeV$^3$ to 0.08 GeV$^3$ \cite {rm4,Lb2}. Then, for each $m_D$, we get a value of $\alpha_{seff}$ corresponding to a value of $\kappa$. Solving the discretized Eqs. (12,13), which become an eigenvalue equation, we obtain the numerical results for $\tilde{f}_1(p_t)$ and $\tilde{f}_2(p_t)$, which depend on two parameters, $m_D$ and $\kappa$.

 Using Lorentz symmetry and discrete C, P, T symmetries, one can show that the following matrix elements of the $\Lambda_b\rightarrow \Lambda$ transition can be parametrized by twelve independent form factors \cite{cm},

\begin{eqnarray}
&&\langle\Lambda(P',s')\arrowvert \bar{s}\gamma_{\mu}b\arrowvert\Lambda_b(P,s)\rangle =\nonumber\\
&&\bar{u}_{\Lambda}(P',s')(g_1\gamma^\mu+ ig_2\sigma_{\mu\nu}p^{\nu}+g_3p_\mu)u_{\Lambda_b}(P,s),\nonumber\\
&&\langle\Lambda(P',s')\arrowvert \bar{s}\gamma_{\mu}\gamma_{5}b\arrowvert\Lambda_b(P,s)\rangle=\nonumber\\
&&\bar{u}_{\Lambda}(P',s')(t_1\gamma^\mu+it_2\sigma_{\mu\nu}p^{\nu}+t_3p^\mu)\gamma_5u_{\Lambda_b}(P,s),\nonumber\\
&&\langle\Lambda(P',s')\arrowvert \bar{s}i\sigma^{\mu\nu}q^{\nu}b\arrowvert\Lambda_b(P,s)\rangle=\nonumber\\
&&\bar{u}_{\Lambda}(P',s')(g^T_1\gamma^\mu+ig^T_2\sigma_{\mu\nu}q^{\nu}+g^T_3q^\mu)u_{\Lambda_b}(P,s),\nonumber\\
&&\langle\Lambda(P',s')\arrowvert \bar{s}i\sigma^{\mu\nu}\gamma_5q^{\nu}b\arrowvert\Lambda_b(P,s)\rangle=\nonumber\\
&&\bar{u}_{\Lambda}(P',s')(t^T_1\gamma^\mu+it^T_2\sigma_{\mu\nu}q^{\nu}+t^T_3q^\mu)\gamma_5u_{\Lambda_b}(P,s),
\end{eqnarray}

\noindent where $q=P-P'$ is the momentum transfer, and $g_i$, $t_i$, $g_i^T$, $t_i^{T}$ ($i=1,2$ and 3) are various form factors which are Lorentz scalar functions of $q^2$. The most general form for the matrix elements in Eq. (16) consistent with the spin symmetry on the $b$ quark in the limit
$m_b\rightarrow \infty $ is

\begin{eqnarray}
  \langle\Lambda(P',s')\arrowvert \bar{s}\Gamma_{\mu} b\arrowvert && \Lambda_b(v,s)\rangle =\bar{u}_{\Lambda}(P',s')(F_{1}+F_2\slashed{v})\nonumber\\
  &&\Gamma^{\mu}u_{\Lambda_b}(v,s),
\end{eqnarray}

\noindent where $\Gamma_{\mu}$ represent $\gamma_{\mu}$, $\gamma_{\mu}\gamma_5$, $\sigma_{\mu\nu}q^{\nu}$, and $\sigma_{\mu\nu}\gamma_5q^{\nu}$. $F_i$ ($i=1, 2$) can be expressed as functions solely of $v\cdot P'$, which is the energy of the $\Lambda$ baryon in the $\Lambda_b$ rest frame. Comparing Eq. (16) with Eq. (17), we obtain the following relations:

\begin{eqnarray}
 & & g_1~=~t_1~=~g^T_2~=~t^T_2~=~\bigg(F_1+\frac{M_{\Lambda}}{M_{\Lambda_{b}}}F_2\bigg),\nonumber\\
 & & g_2~=~t_2~=g_3~=~t_3~=~\frac{1}{M_{\Lambda_{b}}}F_2, \nonumber\\
 & & g^T_3~=~-\frac{F_2}{M_{\Lambda_{b}}}(M_{\Lambda_b}-M_{\Lambda}),\nonumber \\
 & & t^T_3~=~\frac{F_2}{M_{\Lambda_{b}}}(M_{\Lambda_b}+M_{\Lambda}), \nonumber\\
 & & g_1^T~=~ t^T_1~=~\frac{F_2}{M_{\Lambda_{b}}}q^2.
\end{eqnarray}

The BS wave function of $\Lambda_b$ was given in previous works and has the form $\chi_P^{\Lambda_b}(p)=\phi^{\Lambda_b}(p)u_{\Lambda_b}(v,s)$, where $\phi^{\Lambda_b}(p)$ is the scalar BS wave functions \cite{Lb,Lb2}. The transition matrix for $\Lambda_b\rightarrow \Lambda$ can be expressed in terms of the BS wave function of $\Lambda_b$ and the $s(ud)_{0,0}$ component of $\Lambda$, $\chi^{\Lambda}_{P'}(p')$,

\begin{eqnarray}
  \langle\Lambda(P',s')\arrowvert\bar{s}\Gamma_{\mu}\arrowvert\Lambda_b(P,s)\rangle & & =\int\frac{d^4p}{(2\pi)^4}\times\nonumber\\
 & &\bar{\chi}_{P'}^{\Lambda}(p')\Gamma_{\mu}\chi_P^{\Lambda_b}(p)S^{-1}_D(p_2).
\end{eqnarray}

\noindent From Eqs. (3) and (4) we can see that the Clebsh-Gordan coefficient of the $s(ud)_{0,0}$ configuration is $1/\sqrt{3}$. Substituting the BS wave functions of $\Lambda_b$ and the $s(ud)_{0,0}$ system into Eq. (19) and integrating out the longitudinal momentum $p_l$, we obtain the following forms for $F_1$ and $F_2$:

\begin{eqnarray}
F_1 &= &\frac{1}{\sqrt{3}}\int\frac{d^3p_t}{(2\pi)^3}\tilde{\phi}^{\Lambda_b}(p_t)\int\frac{d^3k_t}{(2\pi)^3}\nonumber\\
& &
\bigg\{-\frac{1}{2\omega'_s(M_\Lambda-\omega\omega_D-\omega'_s-\sqrt{\omega^2-1}\cos{\theta}p_t)} \nonumber \\
& &\big\{\big[(\tilde{V}_1(p'_t-k_t)+2\omega_D\tilde{V}_2(p'_t-k_t)F(Q^2))(m_s+\omega'_s)\nonumber \\
& & -(p'_t\cdot k_t+p'^2_t)\tilde{V}_2(p'_t-k_t)F(Q^2)\big]\tilde{f}^{\Lambda}_1(k_t) \nonumber \\
& & +\big[(\tilde{V}_1(p'_t-k_t)-2\omega{_D}\tilde{V}_2(p'_t-k_t)F(Q^2))\nonumber \\
& &-(m_s+\omega'_s) (p'_t\cdot k_t+p'^2_t)\tilde{V}_2F(Q^2)\big]\tilde{f}^{\Lambda}_2(k_t)\big\}\nonumber \\
& & +\frac{\omega}{1-\omega^2} \frac{1}{2\omega'_s(M_\Lambda-\omega\omega_D-\omega'_s-\sqrt{\omega^2-1}\cos{\theta}p_t)}\nonumber \\
& & v\cdot p'_t \times\big\{\big[(\tilde{V}_1(p'_t-k_t)+2\omega_D\tilde{V}_2(p'_t-k_t)F(Q^2))\nonumber \\
& & -(m_s-\omega'_s)(\tilde{V}_1(p'_t-k_t)\nonumber\\
& &+\frac{p'_t\cdot k_t}{p'^2_t}\tilde{V}_2(p'_t-k_t)F(Q^2))\big]\tilde{f}^{\Lambda}_1(k_t) \nonumber \\
& & +\big[(m_s-\omega'_s)(\tilde{V}_1(p'_t-k_t)\nonumber\\
& &-2\omega_D \tilde{V}_2(p'_t-k_t)F(Q^2))\frac{p'_t\cdot k_t}{p'^2_t}\nonumber\\
& & -(p'_t\cdot k_t+p'^2_t)\tilde{V}_2(p'_t-k_t)F(Q^2)\big]\tilde{f}^{\Lambda}_2(k_t)\big\}\bigg\},
\end{eqnarray}

\begin{eqnarray}
F_2&=&\frac{1}{\sqrt{3}}\int\frac{d^3p_t}{(2\pi)^3}\tilde\phi^{\Lambda_b}(p_t)\int\frac{d^3k_t}{(2\pi)^3}\nonumber\\
& & -\frac{1}{1-\omega^2}\frac{1}{2\omega'_s(M_\Lambda-\omega\omega_D-\omega'_s-\sqrt{\omega^2-1}\cos{\theta}p_t)}\nonumber \\
& &v\cdot p'_t \times\bigg\{\big[(\tilde{V}_1(p'_t-k_t)+2\omega_D \tilde{V}_2(p'_t-k_t)F(Q^2))\nonumber\\
& &-(m_s-\omega'_s)(\tilde{V}_1(p'_t-k_t)+\frac{p'_t\cdot k_t}{p'^2_t}\tilde{V}_2F(Q^2))\big]\tilde{f}^{\Lambda}_1(k_t) \nonumber \\
& & +\big[(m_s-\omega'_s)(\tilde{V}_1(p'_t-k_t)-2\omega_D \tilde{V}_2(p'_t-k_t)F(Q^2))\nonumber\\
& &\frac{p'_t\cdot k_t}{p'^2_t}-(p'_t\cdot k_t+p'^2_t)\tilde{V}_2(p'_t-k_t)F(Q^2)\big]\tilde{f}^{\Lambda}_2(k_t)\bigg\},\nonumber\\
& &
\end{eqnarray}

\noindent where $\omega_s=\sqrt{m_s^2-p^{2}_t}$, $\omega'_s=\sqrt{m_s^2-p^{'2}_t}$, $p'_t~(=p'-p'_l\cdot v)$ and $p'_l~(=p'\cdot v)$ are the transverse and longitudinal relative momenta along the momentum of $\Lambda$, respectively, $\omega=v\cdot v'$ ($v$ and $v'$ are the velocities of the $\Lambda_b$ and $\Lambda$, respectively) is the invariant velocity transfer, $\theta$ is the angle between $p_t$ and $v'_t$. All the form factors are functions of the invariant velocity transfer, $ \omega = \frac{m_{\Lambda_b}^2+m_{\Lambda}^2-q^2}{2m_{\Lambda_b} m_{\Lambda}}$, therefore, the minimum and maximum values of $\omega$ are 1 and $\frac{m_{\Lambda}^2+m_{\Lambda_b}^2}{2m_{\Lambda_b} m_{\Lambda}}$, respectively.
In our calculation, we take $m_s=0.45$ GeV, $M_\Lambda=1.116$ GeV, $M_{\Lambda_b}=5.62$ GeV. Then one can find $\omega$ ranges from 1 to 2.62. Substituting $F_1$ and $F_2$ into Eq. (18) we will get the numerical results of $g_i$, $t_i$, $g_i^T$ and $t_i^{T}$ as functions of $\omega$. The plots of $g_1$, $g_2$, $g^T_1$, $g^T_3$, and $t_3^T$ are shown in Figs 1 and 2. Other form factors can be obtained from Eq. (18) straightforwardly.

\begin{figure}[htb]
\begin{center}\scalebox{0.78}{
\includegraphics [-5,390][290,600]{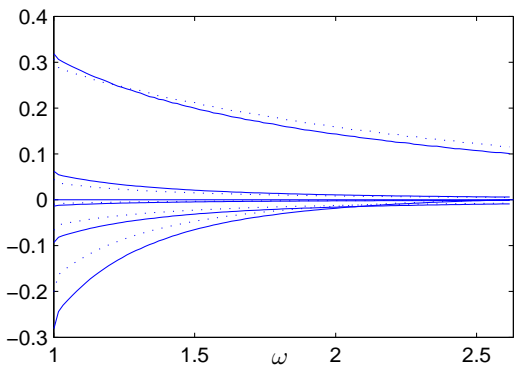}}
\caption{$\Lambda_b\rightarrow\Lambda$ form factors as functions of $\omega$. The upper (lower) line in the upper plane corresponds to $g_1$ ($g_3^T$). The upper, middle, and lower lines in the lower plane correspond to $g_2$, $g_1^T$, and $t_3^T$, respectively. The solid and dotted lines correspond to $m_D=0.7$ GeV and $m_D=0.8$ GeV, respectively, when $\kappa=0.05$GeV$^3$.  }
\end{center}
\end{figure}

\begin{figure}[htb]
\scalebox{0.78}{
\includegraphics [-5,400][290,600] {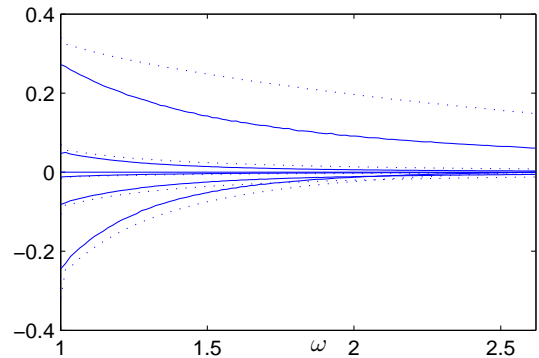}}
\caption{$\Lambda_b\rightarrow\Lambda$ form factors as functions of $\omega$. The upper (lower) line in the upper plane corresponds to $g_1$ ($g_3^T$). The upper, middle, and lower lines in the lower plane correspond to $g_2$, $g_1^T$, and $t_3^T$, respectively. The solid and dotted lines correspond to $\kappa=0.02~$GeV$^3$ and 0.08 $ $GeV$^3$, respectively, when $m_D=0.75$ GeV.  }
\end{figure}

In a similar way, we obtain the form factors for $\Lambda_b\rightarrow p$ replacing $m_s$ by $m_u$ and $M_{\Lambda}$ by $M_p$. $\omega$ for $\Lambda_b\rightarrow p$ ranges from 1 to 3.08, and the the Clebsch-Gorden coefficient of the $u(ud)_{0,0}$ configuration is $1/\sqrt{2}$. The numerical results for $g_1$, $g_2$, $t^T_1$, $t_3^T$, and $g^T_3$ for $\Lambda_b\rightarrow p$ are plotted in Fig. 3 and 4.

\begin{figure}[htb]
\begin{center}\scalebox{0.8}{
\includegraphics [6,400][280,600] {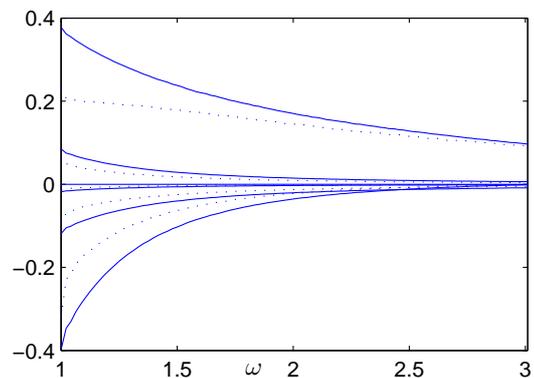}}
\caption{$\Lambda_b\rightarrow p$ form factors as functions of $\omega$. The upper (lower) line in the upper plane corresponds to $g_1$ ($g_3^T$). The upper, middle, and lower lines in the lower plane correspond to $g_2$, $g_1^T$, and $t_3^T$, respectively. The solid and dotted lines correspond to $m_D=0.7$ GeV and $m_D=0.8$ GeV, respectively, when $\kappa=0.05~$ GeV $^3$. }
\end{center}
\end{figure}
\begin{figure}[htb]
\begin{center}\scalebox{0.8}{
\includegraphics [6,400][280,600] {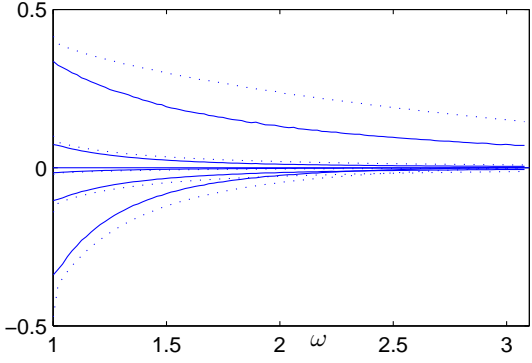}}
\caption{$\Lambda_b\rightarrow p$ form factors as functions of $\omega$. The upper (lower) line in the upper plane corresponds to $g_1$ ($g_3^T$). The upper, middle, and lower lines in the lower plane correspond to $g_2$, $g_1^T$, and $t_3^T$, respectively. The solid and dotted lines correspond to $\kappa=0.02~$GeV$^3$ and 0.08  GeV$^3$, respectively, when $m_D=0.75$ GeV. }
\end{center}
\end{figure}

From Figs. 1-4, we can see that the magnitudes of form factors decrease as $\omega$ increases. This is because the overlap integrals of BS wave functions decrease with the increase of $\omega$. The numerical results of these functions will be used to calculate the decay widths of $\Lambda_b\rightarrow\Lambda l^+l^-$ and $\Lambda_b\rightarrow p l\bar{\nu}$ in next sections.

\section*{III. $\Lambda_b \rightarrow\Lambda l^{+} l^{-}$ Decay in the Standard Model}
We first apply our results for the $\Lambda_b\rightarrow\Lambda$ form factors to calculate the differential fraction for the decay $\Lambda_b\rightarrow\Lambda l^+l^-$ $(l=e, \mu, \tau)$ in the SM. This process is loop-suppressed, and hence potentially sensitive to new physics beyond the SM. At the quark level, $\Lambda_b\rightarrow\Lambda l^+l^-$ is described by $b\rightarrow s l^+l^-$ transition. The effective Hamiltonian describing the electroweak penguin and weak box diagrams related to this transition is given by \cite{SM5,KA}

\begin{eqnarray}
\mathcal{H}&=&\frac{G_F\alpha}{\sqrt{2}\pi}\lambda_t\bigg[C^{eff}_9(\bar{s}\gamma_{\mu}P_Lb)+C_{10}(\bar{s}\gamma_{\mu}P_L b)(\bar{l}\gamma^{\mu}\gamma_5l)\nonumber\\
& &-2C^{eff}_{7}m_b\bigg(\bar{s}i\sigma_{\mu\nu}\frac{q^{\mu}}{q^2}P_R\bigg)(\bar{l}\gamma_{\mu}l)\bigg],
\end{eqnarray}

where $G_F$ is the Fermi coupling constant, $\alpha$ is the electromagnetic coupling constant, $$\lambda_t=V_{tb}V^*_{ts},$$ $$P_{R,L}=\frac{1}{2}(1\pm \gamma_5),$$ $q$ is the momentum transferred to the lepton pair which is the sum of the momenta of $l^+$ and $l^-$, $C^{eff}_7$, $C^{eff}_9$, and $C_{10}$ are the Wilson coefficients. We note that only the term associated with the Wilson coefficient $C_{10}$ is independent of the renormalization scale. To find the transition amplitude $\Lambda_b\rightarrow\Lambda$, we need to sandwich this effective Hamiltonian between the initial and final baryon states and calculate the matrix elements $$\langle\Lambda(P',s')\arrowvert \bar{s}\gamma_{\mu}b\arrowvert\Lambda_b(P,s)\rangle, $$ $$\langle\Lambda(P',s')\arrowvert \bar{s}\gamma_{\mu}\gamma_{5}b\arrowvert\Lambda_b(P,s)\rangle,$$ $$\langle\Lambda(P',s')\arrowvert \bar{s}i\sigma^{\mu\nu}q^{\nu}b\arrowvert\Lambda_b(P,s)\rangle,$$ and $$\langle\Lambda(P',s')\arrowvert \bar{s}i\sigma^{\mu\nu}\gamma_5q^{\nu}b\arrowvert\Lambda_b(P,s)\rangle.$$ These matrix elements are expressed in terms of the form factors  obtained from Eqs. (16-21) in Section II. Then, the matrix element of the decay $\Lambda_b\rightarrow\Lambda l^+l^-$ can be written as \cite{SM5,KA}

\begin{eqnarray}
\mathcal{M}(\Lambda_b & &\rightarrow \Lambda l^{+} l^{-})=\frac{G_F}{2\sqrt{2}\pi}\times \lambda_t\big[\bar{l}\gamma_{\mu}l\{\bar{u}_{\Lambda}[\gamma_{\mu}(A_1(1+\gamma_5)\nonumber\\
& &+B_1(1-\gamma_5))+i\sigma^{\mu\nu}p_{\nu}(A_2(1+\gamma_5)\nonumber\\
& &+B_2(1-\gamma_5))]u_{\Lambda_b}\}\nonumber\\
& &+\bar{l}\gamma_{\mu}\gamma_5l\{\bar{u}_{\Lambda}[\gamma^{\mu}(D_1(1+\gamma_5)\nonumber\\
& &+E_1(1-\gamma_5))+i\sigma^{\mu\nu}p_{\nu}(D_2(1+\gamma_5)+E_2(1-\gamma_5))\nonumber\\
& &+p^{\mu}(D_3(1+\gamma_5)+E_3(1-\gamma_5))]u_{\Lambda_b}\}\big],
\end{eqnarray}

where the parameters $A_i$, $B_i$ and $D_j$, $E_j$ ($i=1,2$ and $j=1,2,3$) are defined as

\begin{eqnarray}
&&A_i=\frac{1}{2}\bigg\{C^{eff}_{9}(g_i-t_i)-\frac{2C^{eff}_7}{p^2}(g_i^T+t_i^T)\bigg\},\nonumber\\
& &B_i=\frac{1}{2}\bigg\{C^{eff}_{9}(g_i+t_i)-\frac{2C^{eff}_7}{p^2}(g_i^T-t_i^T)\bigg\},\nonumber\\
& &D_j=\frac{1}{2}C_{10}(g_j-t_j),\nonumber\\
& &E_j=\frac{1}{2}C_{10}(g_j+t_j).
\end{eqnarray}

The final task is to calculate the decay rate of $\Lambda\rightarrow\Lambda l^+l^-$ in the whole physical region, $4m^2_l\leq q^2\leq (m_{\Lambda_b}-m_{\Lambda})^2$. The differential decay rate is obtained as \cite{SM5,KA}

\begin{eqnarray}
\frac{d\Gamma}{dq^2}=\frac{G^2_F\alpha^2}{8192\pi^5m_{\Lambda_b}}& &|V_{tb}V^*_{ts}|^2v_l\sqrt{\lambda(1,r,s)}[\mathcal{T}_0(s)\nonumber\\
& &+\frac{1}{3}\mathcal{T}_2(s)],
\end{eqnarray}

where $$s=q^2/m^2_{\Lambda_b},$$ $$r=m_{\Lambda}/m_{\Lambda_b},$$ $$\lambda(1,r,s)=1+r^2+s^2-2r-2s-2rs,$$ and $$v_l=\sqrt{1-\frac{4m^2_l}{q^2}}$$ is the lepton velocity. The functions $\mathcal{T}_0(s)$ and $\mathcal{T}_2(s)$ are given as \cite{SM5,KA}

\begin{eqnarray}
  \mathcal{T}_0(s)&&=32m^2_l m^4_{\Lambda_b}s(1+r-s)(|D_3|^2+|E_3|^2)\nonumber\\
& &+64m^2_lm^3_{\Lambda_b}(1-r-s)Re(D^*_1E_3+D_3E^*_1)\nonumber\\
& &+64m^2_{\Lambda_b}\sqrt{r}(6m^2_l-M^2_{\Lambda_b}s)Re(D_1^*E_1)\nonumber\\
& & +64m^2_lm^3_{\Lambda}\sqrt{r}\big(2m_{\Lambda_b}s Re(D^*_3E_3)\nonumber\\
& &+(1-r+s)Re(D^*_1D_3+E^*_1E_3)\big)\nonumber\\
& &+32m^2_{\Lambda}(2m^2_l+m^2_{\Lambda}s)\{(1-r+s)m_{\Lambda_b}\sqrt{r}Re(A^*_1A_2\nonumber\\
& &+B^*_1B_2)-m_{\Lambda_b}(1-r-s)Re(A^*_1B_2+A^*_2B_1)\nonumber\\
& & -2\sqrt{r}\big(Re(A^*_1B_1)+m^2_{\Lambda}s Re(A^*_2B_2)\big)\nonumber\\
& &+8m^2_{\Lambda_b}[4m^2_l(1+r-s)+m^2_{\Lambda_b}((1+r)^2-\nonumber\\
& &s^2)](|A_1|^2+|B_1|^2)+8m^4_{\Lambda_b}\{4m^2_l[\lambda+(1+r-s)s]\nonumber\\
& &+m^2_{\Lambda_b}s[(1-r)^2-s^2]\}(|A_2|^2+|B_2|^2)\nonumber\\
& &-8m^2_{\Lambda_b}\{4m^2_l(1+r-s)-m_{\Lambda_b}[(1-r)^2-s^2]\}\nonumber\\
& & (|D_1|^2+|E_1|^2) +8m^5_{\Lambda_b}sv^2\{-8m_{\Lambda_b}s\sqrt{r}Re(D^*_2E_2)\nonumber\\
& &+4(1-r+s)\sqrt{r}Re(D^*_1D_2+E^*_1E_2) -4(1-r-s)\nonumber\\
& &Re(D^*_1E_2+D^*_2E_1)+m_{\Lambda_b}[(1-r)^2-s^2]\nonumber\\
& &(|D_2|^2+|E_2|^2)\},
\end{eqnarray}

and

\begin{eqnarray}
\mathcal{T}_{2}(s)&=&-8m^4_{\Lambda_b}v_l^2\lambda(|A_1|^2+|B_1|^2+|C_1|^2+|D_1|^2)\nonumber\\
& &+8m^6_{\Lambda_b}s v_l^2\lambda(|A_2|^2+|B_2|^2+|C_2|^2+|D_2|^2).\nonumber\\
& &
\end{eqnarray}

$\omega=(m^2_{\Lambda_b}+m^2_{\Lambda}-q^2)/2m_{\Lambda_b}m_{\Lambda}$, so $\omega$ ranges from 1 to $(m^2_{\Lambda}+m^2_{\Lambda_b}-4m^2_l)/2m_{\Lambda}m_{\Lambda_b}$. The differential decay rate expressed in terms of $\omega$ has the following form,

\begin{eqnarray}
\frac{d\Gamma}{d(2m_{\Lambda} m_{\Lambda_b}\omega)}&=&\frac{G^2_F\alpha^2m_{\Lambda_b}}{8192\pi^5}|V_{tb}V^*_{ts}|^2v_l\sqrt{\lambda(1,r,s)}[\mathcal{T}_0(s)\nonumber\\
& &+\frac{1}{3}\mathcal{T}_2(s)].
\end{eqnarray}

In our numerical calculations, we use the value of the CKM matrix elements $|V_{tb}V^*_{ts}|=0.041$ and the Wilson coefficients at $\mu=m_b$, $C^{eff}_7=-0.313$, $C^{eff}_9=4.334$ and $C_{10}=-4.669$ \cite{WC,WC1,WC2}. As mentioned before, letting $\kappa$ range from $0.02$ GeV  to $0.08$ GeV and $m_D$ from 0.7 MeV to 0.8 MeV, we have obtained numerical results of the form factors $g(t)_i$, $g^T(t^T)_i$ $(i=1,2,3)$. Using the lifetime of the $\Lambda_b$ baryon, $(1.451\pm0.013)\times10^{-12}s$ \cite{pdg}, and integrating the differential branching ratio (28) over $\omega$ from 1 to $(m^2_{\Lambda}+m^2_{\Lambda_b}-4m^2_l)/2m_{\Lambda}m_{\Lambda_b}$, we obtain the ranges of the branching ratios, which are listed in Table I.

\begin{table*}[htb]
\caption{Values of the branching ratios for $\Lambda_b\rightarrow \Lambda l^+l^-$ in our model and the values from the light-cone QCD sum rules and HQET for different leptons}
\label{sphericcase}
\begin{tabular*}{\textwidth}{@{\extracolsep{\fill}}lrrrl@{}}
\hline\hline
&present work & HQET \cite{cs} & light-cone QCD sum rules\cite{SM5} & Exp. \cite{pdg}  \\
\hline
$Br(\Lambda_b\rightarrow\Lambda e^+e^-)$ & $(1.21\sim2.32)\times 10^{-6} $ &$(2.23\sim 3.34 )\times 10^{-6}$&$(4.6\pm1.6)\times 10^{-6}$\\
$Br(\Lambda_b\rightarrow\Lambda \mu^+\mu^-)$ & $(0.53\sim0.89)\times 10^{-6} $ &$(2.08\sim 3.19 )\times 10^{-6}$&$(4.0\pm1.2)\times 10^{-6}$&$(1.08\pm0.28)\times10^{-6}$\\
$Br(\Lambda_b\rightarrow\Lambda \tau^+ \tau^-)$ & $(0.037\sim0.083)\times 10^{-6}$ &$(0.179\sim 0.276 )\times 10^{-6}$&$(0.8\pm0.3)\times 10^{-6}$\\
\hline
\end{tabular*}
\end{table*}

In Table I, we also present the values of the branching ratios obtained in HQET \cite{cs} and the light-cone QCD sum rules \cite{SM5}. It can be seen from the Table I that, as is excepted, the branching ratios decrease when $l$ goes from the $e$ to $\tau$ \cite{WC2}. We can also see that our result on the branching ratio for $\Lambda_b\rightarrow\Lambda \mu^+\mu^-$ is about 1/4 of that predicted by HQET and about 1/6 of that given by light-cone QCD sum rules. Our result is consistent with the experimental data and the other two are not. Since $10^{10}$$\sim$$10^{11}$ $\Lambda_b\bar{\Lambda}_b$ pairs are expected to be produced per year at LHCb, the results presented in Table I indicate that the detection possibility of $\Lambda_b\rightarrow\Lambda l^+l^-$ $(l=e,\mu,\tau)$ is quite high \cite{SM5}.

Letting $\kappa$ and $m_D$ vary in their regions we obtain the area of the differential branching ratio for $\Lambda_b\rightarrow\Lambda \mu^+\mu^-$, which is shown in Fig 5 along with recent experimental results from LHCb \cite{LHCb}. The agreement of our results with the experimental data is clear except when the square of the momentum transfer is bigger than 15 GeV$^2$. So there is still room for possible new physics. We also compared our result with that of the first Lattice QCD simulation \cite{SM5,PS} and found that they are consistent with each other. Predictions for $\Lambda_b\rightarrow\Lambda l^+l^-$ when $l=e, \tau$ are shown in Fig. 6 and Fig. 7, respectively, which will be compared with the forthcoming experimental data.

\begin{figure}[htb]
\begin{center}
\scalebox{0.95}{\includegraphics [width=\columnwidth] {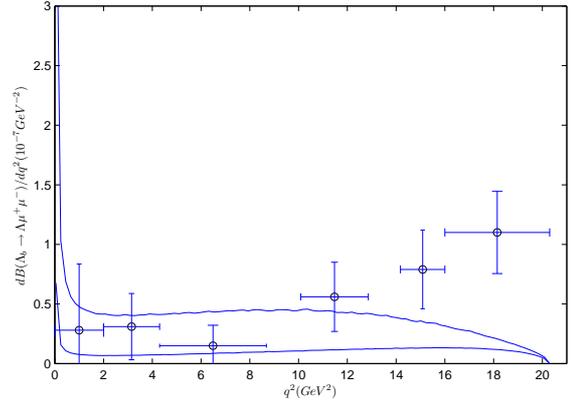}}
\caption{The differential branching ratio for $\Lambda_b\rightarrow\Lambda\mu^+\mu^-$ obtained in our model. The upper (lower) curved solid line corresponds to the upper (lower) boundary of the differential branching ratio as $m_D$ and $\kappa$ vary in their ranges. The experimental data are taken from Ref. [11], with the error bars including systematic and statistical uncertainties.}
\end{center}
\end{figure}

\begin{figure}[htb]
%\centering
\scalebox{0.95}{
\includegraphics [width=\columnwidth] {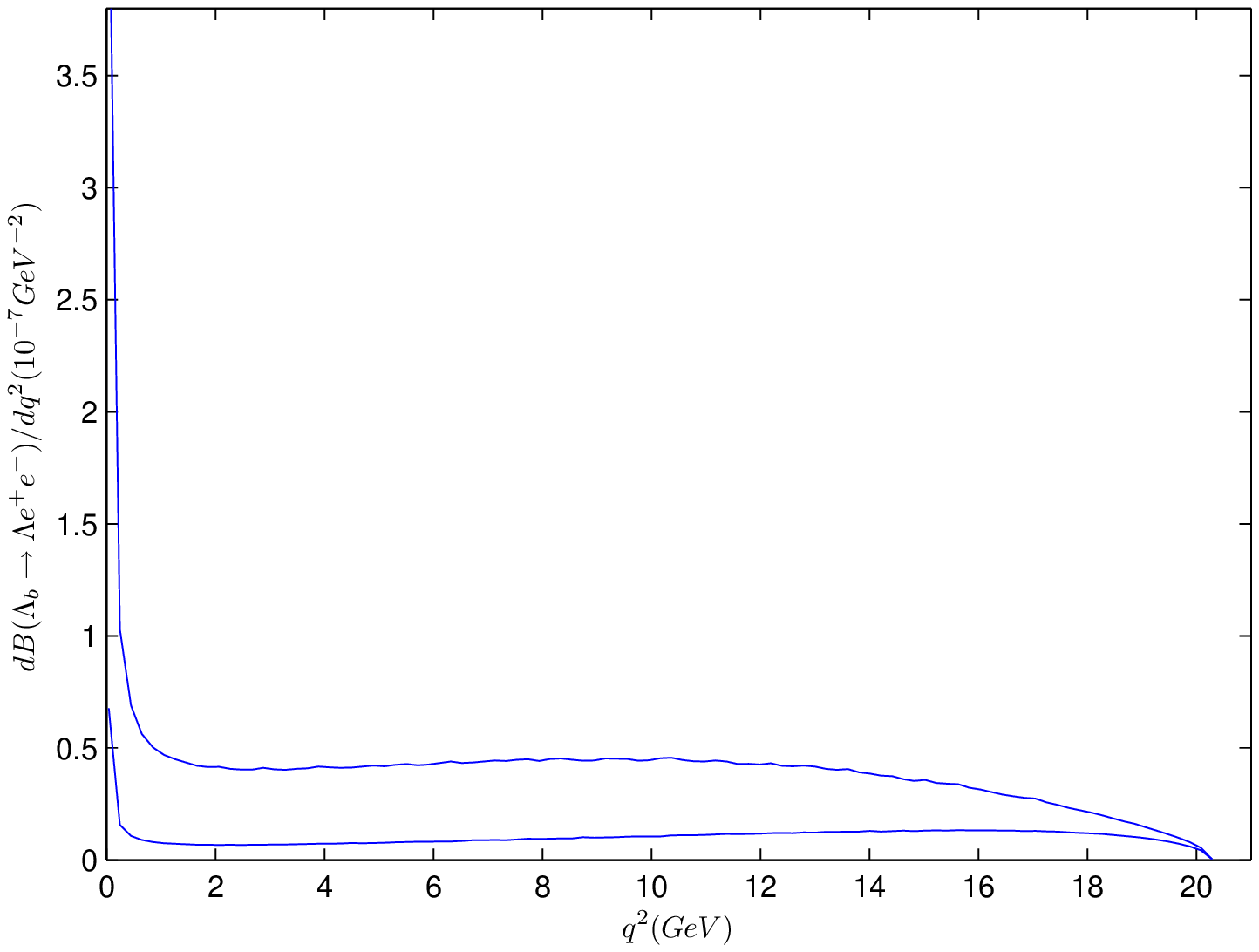}}
\caption{The differential branching ratio for $\Lambda_b\rightarrow\Lambda e^+ e^-$ obtained in our model. The upper (lower) curved solid line corresponds to the upper (lower) boundary of the differential branching ratio. The uncertainties of the boundaries come from the ranges of $m_D$ and $\kappa$.}
\end{figure}

\begin{figure}[htb]
\begin{center}
\scalebox{0.75}{
\includegraphics [60,340][380,600] {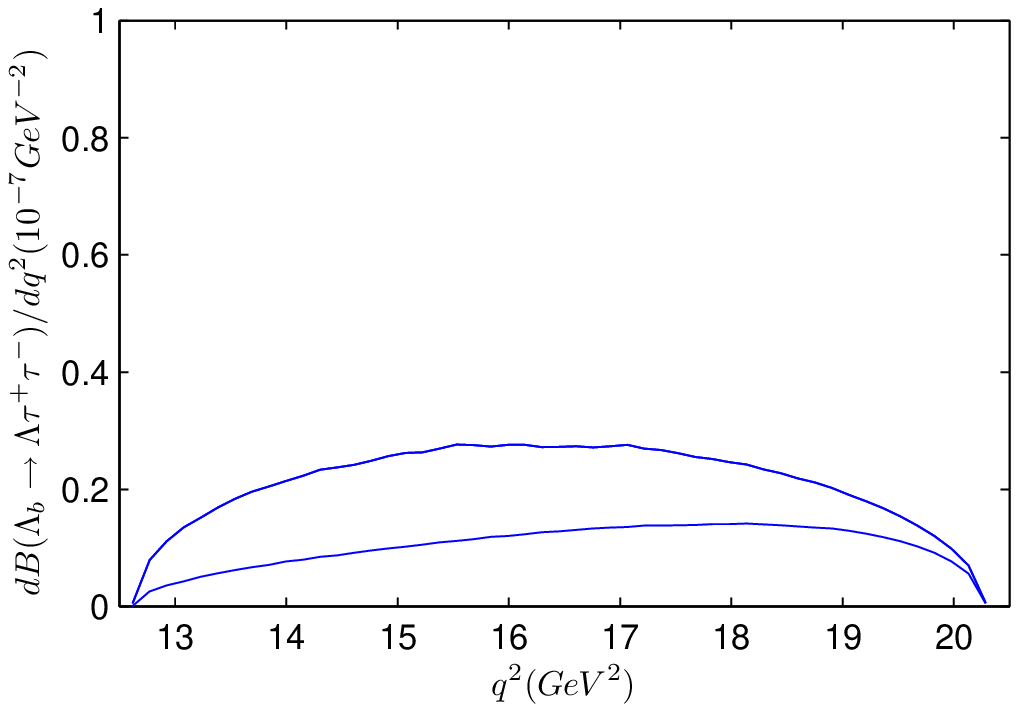}}
\caption{The differential branching ratio for $\Lambda_b\rightarrow\Lambda\tau^+\tau^-$ obtained in our model. The upper (lower) curved solid line corresponds to the upper (lower) boundary of the differential branching ratio. The uncertainties of the boundaries come from the ranges of $m_D$ and $\kappa$.}
\end{center}
\end{figure}

\section*{IV. THE DECAY $\Lambda_b\rightarrow p l\bar{\nu} $}
At the quark level, the decay $\Lambda_b\rightarrow p l\bar{\nu}$ is governed by the tree-level $b\rightarrow u$ transition. The effective Hamiltonian responsible for this transition at the quark level has the form
\begin{eqnarray}
  \mathcal{H}(\Lambda_b\rightarrow p l \bar{\nu})=\frac{G_f}{\sqrt{2}}V_{ub}\bar{u}\gamma_{\mu}(1-\gamma_5)b\bar{l}\gamma^{\mu}(1-\gamma_5)\nu.
\end{eqnarray}
To calculate the amplitude, we need to sandwich the above Hamiltonian between the initial and final states and compute the matrix element $\langle p|\bar{u}\gamma_{\mu}(1-\gamma_5)b|\Lambda_b\rangle$. As discussed in Sect. I, we have obtained the six form factors for this decay amplitude in our previous work \cite{rm4}.

In the next step, we calculate the total decay rate of $\Lambda_b\rightarrow p l\bar{\nu}$ in the whole physical region, $$m^2_l\leq q^2\leq (m_{\Lambda_b}-m_p)^2.$$ The range of $\omega$ for this decay is $$[1,(m^2_{\Lambda_b}+m^2_{\Lambda}-m^2_l)/(2m_{\Lambda_b}m_{\Lambda})]$$ The decay width is given by the following expression:
\begin{eqnarray}
  \Gamma(\Lambda_b\rightarrow p l \bar{\nu})&=&\frac{G^2_f}{384\pi^3 m^3_{\Lambda_b}}|V_{ub}|^2\int^{\Delta^2}_{m^2_l}dq^2(1-m^2_l/q^2)^2\nonumber\\
& &\times\sqrt{(\Sigma^2-q^2)(\Delta^2-q^2)}N(q^2 ),
\end{eqnarray}
where

\begin{eqnarray}
N(&&q^2)=g^2_1{q^2}(\Delta^2(4q^2-m^2_l)+2\Sigma^2\Delta^2(1+2m^2_l/q^2)\nonumber\\
& &-(\Sigma^2+2q^2)(2q^2+m^2_l))+g^2_2(q^2)(\Delta^2-q^2)\nonumber\\
& &(2\Sigma^2+q^2)\times(2q^2+m^2_l)+3M^2_{\Lambda_b}g_3^2(q^2)m^2_l(\Sigma^2\nonumber\\
& &-q^2)q^2+6g_1(q^2)g_2(q^2)(\Delta^2-q^2)(2q^2+m^2_l)\Sigma\nonumber\\
& &-6g_1(q^2)g_3{q^2}m^2_l(\Sigma^2-q^2)\Delta+t^2_1(q^2)(\Sigma^2(4q^2-m^2_l)\nonumber\\
& &+2\Sigma^2\Delta^2(1+2m^2_l)/q^2-(\Delta^2+2q^2)(2q^2+m^2_l))\nonumber\\
& &+t^2_2(q^2)(\Sigma^2-q^2)(2\Delta^2+q^2)(2q^2+m^2_l)\nonumber\\
& &+3t^2_3(q^2)m^2_l(\Delta^2-q^2)q^2-6t_1(q^2)t_2(q^2)(\Sigma^2-q^2)\nonumber\\
& &(2q^2+m^2_l)\Delta+6t_1(q^2)g_3*(q^2)m^2_l(\Delta^2-q^2)\Sigma,
\end{eqnarray}

 and $\Delta=m_{\Lambda_b}-m_p$, $\Sigma=m_{\Lambda_b}+m_p$. The numerical results are listed in Table II, together with theoretical results from other models.

 \begin{table*}[htb]
\caption{Values of the decay rates (in units $|V_{ub}|^2$ $ps^{-1}$) of the  transitions $\Lambda_b\rightarrow p l\bar{\nu}$ and comparision with other methods}
\label{sphericcase}
\begin{tabular*}{\textwidth}{@{\extracolsep{\fill}}lrl@{}}
\hline\hline
&present work & Other models  \\
\hline
$\Lambda_b\rightarrow p\mu^-\bar{\nu}_{\mu}$&$ 3.12\sim7.06$  &$250\pm{85}$\cite{KA};$235\pm{85}$\cite{KA};$477\pm{175}$\cite{KA}; $3.84\pm{1.25}$\cite{KA}; 13.3\cite{TG}; 6.48\cite{DA};4.55\cite{MP};7.55\cite{MP};\\
$\Lambda_b\rightarrow p e^-\bar{\nu}_{e}$&$ 2.09\sim7.40$  &6.48\cite{DA};4.55\cite{MP};7.55\cite{MP}; 13.3\cite{TG}; $250\pm85$\cite{KA}; $235\pm{85}$\cite{KA}; $478\pm{175}$\cite{KA}; $3.76\pm{1.20}$\cite{KA}; \\
$\Lambda_b\rightarrow p \tau^-\bar{\nu}_{\tau}$&$ 1.35\sim4.10$  &4.01\cite{MP}; 6.55\cite{MP}; $312\pm{105}$\cite{KA}; $208\pm{70}$\cite{KA}; $646\pm215$\cite{KA}; $1.93\pm0.70$\cite{KA}; 9.6\cite{TG}; \\
\hline
\end{tabular*}
\end{table*}
The four results of Ref. [56] in each line in the table refer to those from QCD sum rules, lattice QCD, QCD sum rules in the heavy quark limit, and lattice QCD in the heavy quark limit in order. The decay rates were calculated in the covariant quark model \cite{TG}, $SU(3)$ symmetry quark model \cite{DA} and HONR and HOSR constituent quark models (HONR and HOSR refer to harmonic oscillator nonrelativistic and harmonic oscillator semirelativistic  constituent quark models, respectively) \cite{MP}. We compare our results for the rates [in the units of $|V_{ub}|^2$ $ps^{-1}$] with the predictions of other phenomenological methods. From the table, it is clear that our results are of the same order as those of lattice QCD in the heavy quark limit \cite{KA} and those from Refs. [58-59]. However, our results disagree (up to two orders of magnitude) with those of Ref. [57] and those obtained from QCD sum rules, lattice QCD, and QCD sum rules in the heavy quark limit \cite{KA}.

\section*{V. SUMMARY AND DISCUSSION }
Theoretical studies of the rare baryon decay of $\Lambda_b\rightarrow\Lambda l^+l^-$ require knowledge of the hadronic matrix element $\langle\Lambda|\bar{s}\Gamma b|\Lambda_b\rangle$ which involves nonperturbative QCD effects. At the leading order in HQET, this matrix element is described by two independent form factors, which are determinated by the wave functions of the initial and final baryons. We calculate these two form factors in the BS equation approach in the quark-diquark model. Consequently, we obtain all the twelve form factors resposible for the decay $\Lambda_b\rightarrow\Lambda l^+l^-$, which depend on the two parameters, $m_D$ and $\kappa$, in our model. Then, we obtain the total and the differential branching ratios of the decay $\Lambda_b\rightarrow \Lambda l^+l^-$. We also compare our results with those of other approaches and the experimental data from LHC. We find that our result on the total branching ratio of $\Lambda_b\rightarrow\Lambda \mu^+\mu^-$ is consistent with the experimental data but those of the light-cone QCD sum rules and HQET methods are not. We also obtain the area of the differential branching ratio for $\Lambda_b\rightarrow\Lambda \mu^+\mu^-$, which is consistent with those of the first lattice QCD simulation and with the experimental data except when the square of momentum transfer is bigger than 15 GeV$^2$. This indicates there is still room for possible new physics effects. Furthermore, we give the decay rates of $\Lambda_b\rightarrow p l \bar{\nu}$. We find that our values are the same order as those of Refs. [58-59] and those of lattice QCD in the heavy quark limit in each line in Table II \cite{KA}, but different from those obtained from QCD sum rules, lattice QCD, and QCD sum rules in the heavy quark limit by up to two orders of magnitudes \cite{KA}. The decay $\Lambda_b\rightarrow p l \bar{\nu}$ will likely yield the first determination of the CKM matrix elment $|V_{ub}|$ from $\Lambda_b$ decays at LHC. Our results depend on two parameters in our model, $m_D$ and $\kappa$, which vary in some ranges. This lead to some uncertainties in our results. All our predictions will be tested in the future experiments.

\section*{ACKNOWLEDGMENTS }
This work was supported by the National Natural Science Foundation of China (Project Nos. 11175020 and 11275025).

\end{document}